\documentclass[twocolumn,pra]{revtex4-1}
\usepackage[latin9]{inputenc}
\usepackage{float}
\usepackage{amsmath}
\usepackage{amssymb}
\usepackage{mathtools}
\usepackage{graphicx}
\usepackage{esint}
\usepackage{color}
\usepackage[normalem]{ulem}

\makeatletter
 \@ifundefined{textcolor}{}
 {%
   \definecolor{BLACK}{gray}{0}
   \definecolor{WHITE}{gray}{1}
   \definecolor{RED}{rgb}{1,0,0}
   \definecolor{GREEN}{rgb}{0,1,0}
   \definecolor{BLUE}{rgb}{0,0,1}
   \definecolor{CYAN}{cmyk}{1,0,0,0}
   \definecolor{MAGENTA}{cmyk}{0,1,0,0}
   \definecolor{YELLOW}{cmyk}{0,0,1,0}
 }

\makeatother

\begin{document}

\title{On the adiabatic quantum dynamics of fabricated Ising chains}

\author{Juli\'an Vargas-Grajales and Frederico Brito}

\affiliation{Instituto de F\'isica de S\~ao Carlos, Universidade de S\~ao Paulo, CP
369, 13560-970, S\~ao Carlos, S\~ao Paulo, Brazil}

\begin{abstract}
Physical implementations of quantum computation must be scrutinized
about their reliability under real conditions, in order to be considered
as viable candidates. Among the proposed models, those based on adiabatic
quantum dynamics have shown great potential for solving specific tasks
and have already been successfully implemented using superconducting
devices. In this context, we address the issue of how the fabrication
variations are expected to affect on average the computation results, when only dynamical effects occur. By simulating the dynamics of small-scale systems, it is found a considerable robustness for the computation when analyzing results obtained from ensembles of such machines. In addition, it is also addressed whether conditions for adiabaticity could be taken as quantitative measures of it. From the analysis of four known conditions, it is obtained that none could have such an use.
\end{abstract}
\maketitle

\section{Introduction}

Since the first results demonstrating that a quantum computer could
be used to speed up the solution of some problems in the NP class,
like the prime factorization problem \citep{Shor97}, quantum computation
holds the conjecture (yet to be proven) of being capable of solving
efficiently all problems in NP.

One quantum strategy that has attracted very much attention lately
is known as Adiabatic Quantum Computation (AQC) \citep{Farhi00,Farhi01,LidarRMP},
where the solution of a problem of interest is encoded into the ground
state of a Hamiltonian problem $H_{P}$. Since the direct determination
of the ground state of $H_{P}$ is in general as hard as the original
problem, the AQC strategy avoids such a difficult by exploiting the
adiabatic theorem \citep{MessiahBook}: by performing a convenient
slow evolution of a suitable parametrized Hamiltonian, for which
the ground state determination is an easy task, one can reach with
high fidelity the ground state of $H_{P}$. AQC has been proved universal
and showed to be robust against noise \citep{Childs01,Amin08_2,Amin09_2}.
Indeed, its variant known as Quantum Annealing (QA) \citep{Kadowaki98,Santoro06}
suits the cases of optimization problems where the physical system
is in the presence of a non-zero temperature environment.

Different physical implementations of AQC with few qubits were already
shown, for example, using Nuclear Magnetic Resonance (NMR) techniques
\citep{Steffen03,Xu12} and Rydberg-dressed atoms \citep{Keating13}.
In addition, superconducting qubits \citep{Kaminsky04,You05,Grajcar05,Clarke08}
have also shown great potential due to their easy control and promise
of scalability. Moreover, the first implementations of QA with a system
containing 100's of qubits have been already done using implementations
of superconducting devices \citep{Johnson11,Troyer14,Lanting14}.

A simple mathematical description of the AQC strategy can be captured
by the time dependent Hamiltonian 
\begin{eqnarray}
H(t) & = & (1-\Gamma(t))H_{I}+\Gamma(t)H_{P},\label{eq:aqc Hamiltonian}
\end{eqnarray}
where the envelope function $\Gamma(t)$ interpolates the initial
(easy) $H_{I}$ and final (problem) $H_{P}$ Hamiltonians, if it satisfies
the initial, $\Gamma(t=0)=0$, and final, $\Gamma(t=t_{f})=1$, conditions.
Then, starting in the ground state of $H_{I}$, one can reach the
ground state of $H_{P}$ with high fidelity if an adiabatic evolution
is ensured. The protocol adiabaticity is directly dependent on the
minimum gap 
\begin{eqnarray}
\min_{0\leq t\leq t_{f}}\left[\Delta_{10}(t)\right] & \equiv & \Delta_{min},\label{eq:gap min}
\end{eqnarray}
where $\Delta_{10}(t)\equiv E_{1}(t)-E_{0}(t)$ is the difference
between the instantaneous eigenvalues of $H$ associated with the
ground and first excited instantaneous eigenstates. For the most simple
protocols, like the ones designed with constant time rate interpolation
functions, one finds that the protocol duration that ensures an adiabatic
evolution depends on the minimum gap as $\mathcal{O}(\Delta_{min}^{-2})$
\citep{Farhi00,Amin08_2}. However, it is also known that such a dependence
can be attenuated to reach the scaling $\mathcal{O}(\Delta_{min}^{-1})$
\citep{Schaller06}, if more elaborated protocols, as those using
adaptive interpolation, are used.

Thus, inaccuracies of the system physical
parameters not only can have impact on the fidelity of the computation by considerably
modifying the ground state of the final Hamiltonian,
but it can also compromise the designed dynamics, by altering the conditions for meeting an adiabatic evolution. In this work, we focus on the investigation of the latter effect. For that we simulate dynamics of small-size computations based on Ising chains and determine the fidelity loss as a function of such inaccuracies. In addition, by testing four conditions, we put forward an analysis to verify whether adiabatic conditions could be used as quantitative figures for adiabaticity.

\section{The physical source of noise}

In order to give a clear view of the nature of the noise consider
here, we focus on the implementation of Ising chains using superconducting
flux qubits \citep{Johnson11} and provide a brief discussion about
their Hamiltonian derivation.

The simplest prototype of a flux qubit is comprised of a superconductor
loop interrupted by a Josephson junction (rf-SQUID), whose Hamiltonian
can be written as \citep{Makhlin01}, 
\begin{eqnarray}
H_{rf} & = & \frac{Q^{2}}{2C}+\frac{\left(\Phi-\Phi_{x}\right)^{2}}{2L}-E_{J}\cos\left(\frac{2\pi}{\Phi_{0}}\Phi\right)\!,\label{eq:rf squid}
\end{eqnarray}
where $C$ is the junction capacitance, $Q$ represents the charge
on the capacitor, $\Phi_{x}$ is an applied external magnetic flux
and $L$ is the inductance of the loop. The quantization of the system
is performed by promoting the total magnetic flux threading the loop
$\Phi$ and $Q$ to the status of operators satisfying $[\frac{\Phi}{2\pi},Q]=i\hbar$,
since they are canonically conjugated variables. $E_{J}$ is the Josephson
energy defined as $I_{c}\Phi_{0}/2\pi$, where $I_{c}$ is the junction critical
current and $\Phi_{0}$ is the magnetic flux quantum ($\equiv h/2e$).

A system containing several interacting devices can be constructed
using mutual inductance interaction \citep{Makhlin01,Burkard04},
leading to the interacting Hamiltonian \citep{Johnson11}, 
\begin{equation}
H\!\!=\!\!\sum_{i=1}^{N}\! H_{rf}^{(i)}\!+\!\sum_{i<j=1}^{N}\!\! M^{(ij)}\frac{\left(\!\Phi^{(i)}\!\!-\!\Phi_{x}^{(i)}\!\right)\!\!\left(\!\Phi^{(j)}\!\!-\!\Phi_{x}^{(j)}\!\right)}{L^{(i)}L^{(j)}},\label{eq:set of loops}
\end{equation}
where the superscripts refer to the respective device loop and $M^{(ij)}$
is the mutual inductance between the $i$-th and $j$-th loop.

When considering low-lying energy dynamics under appropriate choice
of physical parameters
\footnote{Cases for which $2\pi L^{(i)}I_{C}^{(i)}/\Phi_{0}>0$
} and applied fluxes $\Phi_{x}^{(i)}$, the multidimensional Hilbert
space associated with each device can be truncated to one spanned
solely by the two lowest eigenenergy states. Such an approximation
leads to their known qubit implementation, and allows one to rewrite
Eq. (\ref{eq:set of loops}) as an effective Hamiltonian of a set
of coupled qubits \citep{Harris10b}, 
\begin{equation}
\begin{multlined}
\mathcal{H}  =  -\frac{1}{2}\sum_{i=1}^{N}\left[2|I_{p}^{i}|\left(\Phi_{x}^{(i)}-\Phi_{0}^{(i)}\right)\sigma_{z}^{(i)}+\Delta\sigma_{x}^{(i)}\right] \\
  +\sum_{i<j=1}^{N}M^{(ij)}|I_{p}^{i}||I_{p}^{j}|\sigma_{z}^{(i)}\sigma_{z}^{(j)},\label{eq:N qubits}
\end{multlined}
\end{equation}
where $\sigma_{x,z}^{(i)}$ are Pauli matrices associated with the
$i$-th device. The parameter $|I_{p}^{i}|$ is the magnitude of the
persistent current flowing through the $i$-th loop, whose control
is performed by the external flux $\Phi_{x}^{(i)}$, $\Phi_{0}^{(i)}$
(may be equal to $\Phi_{0}/2$) is the qubit degeneracy point, and
$\Delta$ represents the tunnelling amplitude, which turns out to
be a constant parameter dependent on all qubit parameters ($C$, $L$
and $E_{J}$). Even though the rf-SQUID can provide a fair implementation
of a qubit, it lacks the level of tunability desired for a qubit,
since $\Delta$ cannot be adjusted \textit{in situ. }Such a difficult
can be overcome if the Josephson junction in the rf-SQUID is replaced
by a small loop interrupted by two other junctions (dc-SQUID). The
presence of the small loop gives an extra knob to control the system's
potential, since one can now apply another external loop to the device.
It is possible to show that, under the right conditions, this new
device will have the same Hamiltonian form of Eq. (\ref{eq:rf squid}),
but with a Josephson energy dependent on the external flux threading
the small loop \citep{Makhlin01}. As an immediate consequence, the
amplitude tunnelling $\Delta$ becomes tunable, leading to much easier
implementations of operations. From here onwards, we consider devices
for which both the persistent current and tunnelling amplitude are
tunable.

As is natural for fabricated devices, the physical parameters determining
the system Hamiltonian Eq. (\ref{eq:N qubits}) present an inherent
spread due to fabrication variations. For superconducting devices,
usual deviations for $C$, $L$ and $I_{c}$ are reported to be circa
5\% \citep{Harris09,Katzgraber16}, which are capable of leading to
noticeable changes of the qubit features, thus degrading the fidelity
of operations performed. Because of that, a synchronization strategy
has already been developed in order to deal with small variations
$(\sim1\%)$ of the device's inductance and critical current \citep{Harris09}.
Indeed, Harris\textit{ et al}. demonstrated theoretically and experimentally
that, by off-setting the applied flux to each device, it is possible
to correct very much the deviations arisen from the variation of those
physical parameters. 

However, despite the success of the method, yet one verifies a discrepancy
of some percent between the corrected persistent current and tunnelling
amplitude and their target values. Recently, it was analyzed \cite{Katzgraber16}
how fluctuations in the qubit couplers and the applied fields would
affect the ground state configuration of the Hamiltonian problem $H_{P}$ of some cases of interest,
demonstrating that such fluctuations can lead to non-negligible perturbations
of the original problem. Here, in addition to that issue, it is taken
the perspective of analyzing the whole AQC, which shall consider
the system time evolution. Accordingly, the main focus of this paper
is to characterize the probability of success when implementing an
AQC processor with ``noisy'' devices, which has the same ground state for the final Hamiltonian.

\section{One-dimensional quantum disordered Ising model}

\begin{figure}[t!]
\begin{centering}
\includegraphics[scale=1]{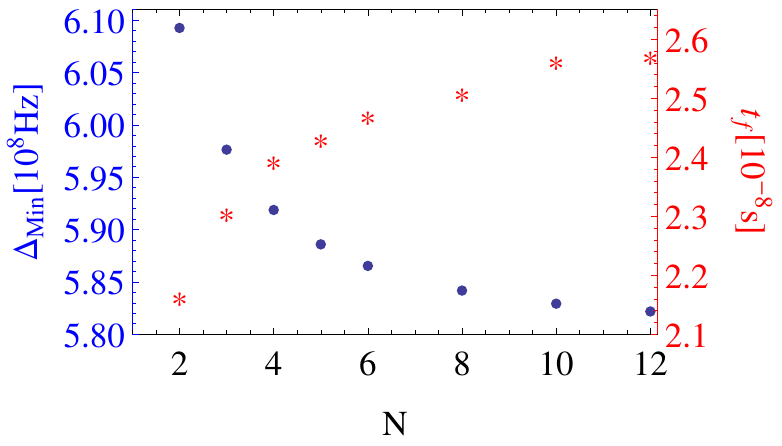} 
\par\end{centering}
\caption{\label{fig:gapminimum} System minimum gap $\Delta_{min}$ (blue dots) and protocol total running time $t_f$ (red stars) as a function of the chain size $N$. The protocol duration $t_{f}$ is chosen for each $N$ such that the adiabaticity is ensured with $\sim99.9975\%$ of probability. For our case study $(\bar{\lambda}_{i}=1,\bar{h}_{i}=5,\bar{J}_{i,i+1}=2.5)$, $\Delta_{min}$ shows a monotonic decreasing behavior with the system size. }
\end{figure}

Based on Hamiltonian (5), we simulate the unitary dynamics determined by
\begin{eqnarray}
{\cal H}_S(t) & = & \Omega(t)H_{I}+\Gamma(t)H_{P},~{\rm with}\label{eq:Hamiltonian}\\
H_{I} & = & -\sum_{i=1}^{N}\frac{\lambda_{i}}{2}\sigma_{x}^{(i)},\nonumber \\
H_{P} & = & -\sum_{i=1}^{N}h_{i}\sigma_{z}^{(i)}-\sum_{i=1}^{N-1}J_{i,i+1}\sigma_{z}^{(i)}\sigma_{z}^{(i+1)},\nonumber 
\end{eqnarray}
considering $\lambda_{i},h_{i}$ and $J_{i,i+1}$ as static random Gaussian
variables with standard deviation $\sigma$, which mean values $(\bar{\lambda}_{i},\bar{h}_{i},\bar{J}_{i,i+1})$
are thought as the ideal implementation of the instance of interest.
Here we choose those mean values and the envelope functions
$(\Omega,\Gamma)$ such that the ground state is always
non-degenerated. In addition to the conditions  $(\Gamma(0)=0,\Omega(0)\neq0)$ and
$(\Gamma(t_{f})\neq0,\Omega(t_{f})=0)$, the protocol is designed such that for the ideal instance
the adiabaticity of the evolution of its initial ground state is satisfied for each chain size $N$ with at least $\sim99.9975\%$ fidelity in the end of the protocol. Furthermore,  the minimum gap $\Delta_{min}$ is found to be monotonically decreasing with the system size (see Fig. \ref{fig:gapminimum}). Once set the protocol for
the ideal instance, we use it for determining the results obtained in an ensemble of 1024 physical realizations of $\lambda_{i},h_{i}$ and $J_{i,i+1}$. 

It is worth of notice that the ideal implementation chosen here also has the ground state of $H_{P}$ insensitive to moderate deviations from its ideal values, i.e. one finds that the instantaneous final ground state is the {\it same} for implementations under those conditions. Therefore, even though the physical realizations may be different, the ground state of their final Hamiltonian give the same (correct) answer to the problem.  Such a feature allows us to certify the source of fidelity loss as a dynamical effect due to solely the break of the adiabaticity for a wide range of $\sigma$ $(\lesssim10\%)$. 

\begin{figure}[t]
\begin{centering}
\hspace{0.8cm}\includegraphics[scale=0.65]{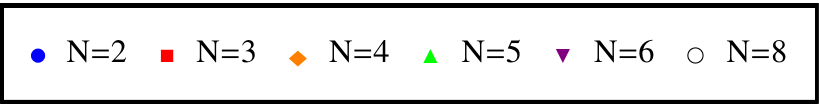}\\
\includegraphics[scale=0.32]{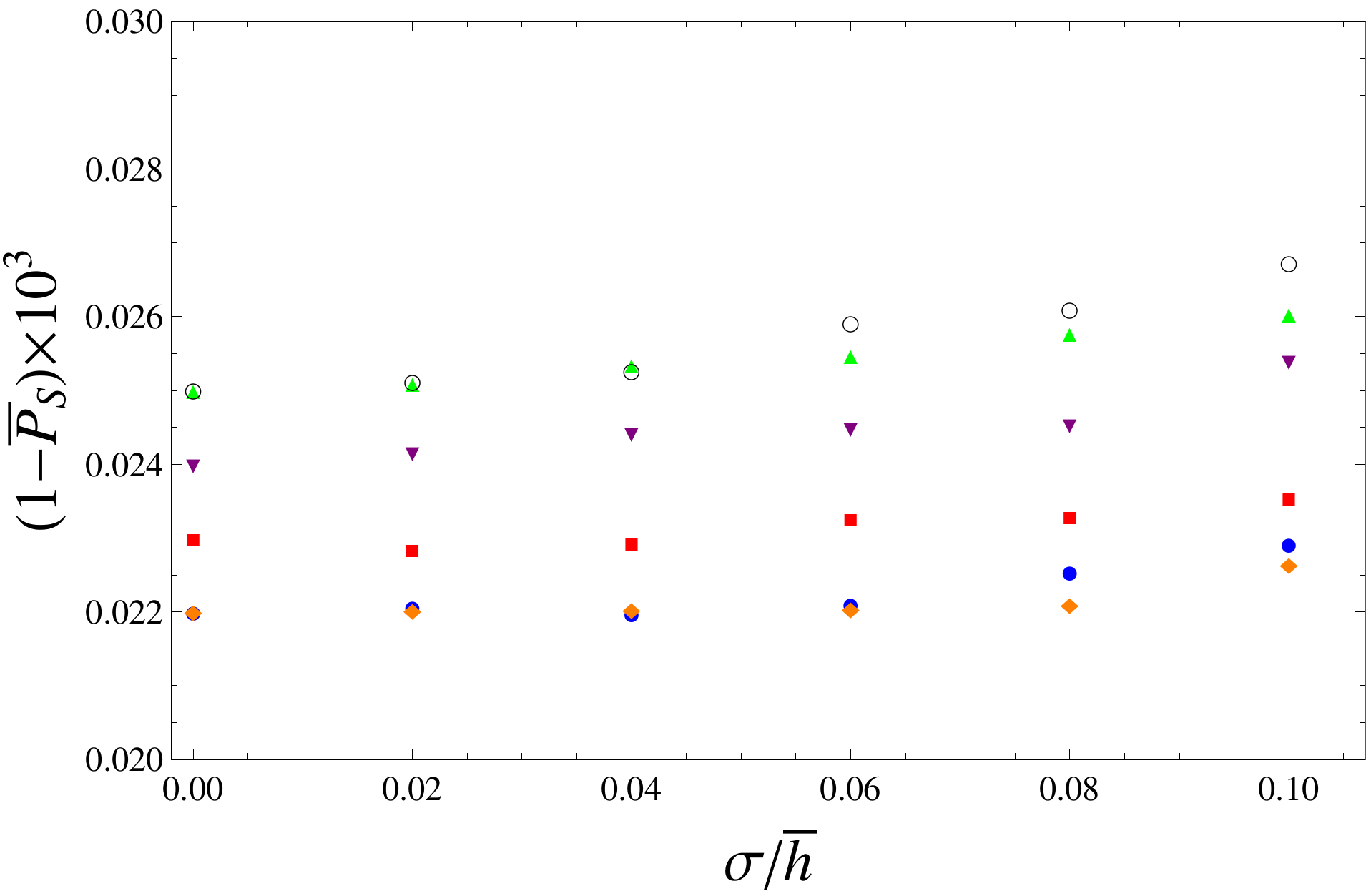}\\
(a)\\
 \includegraphics[scale=0.32]{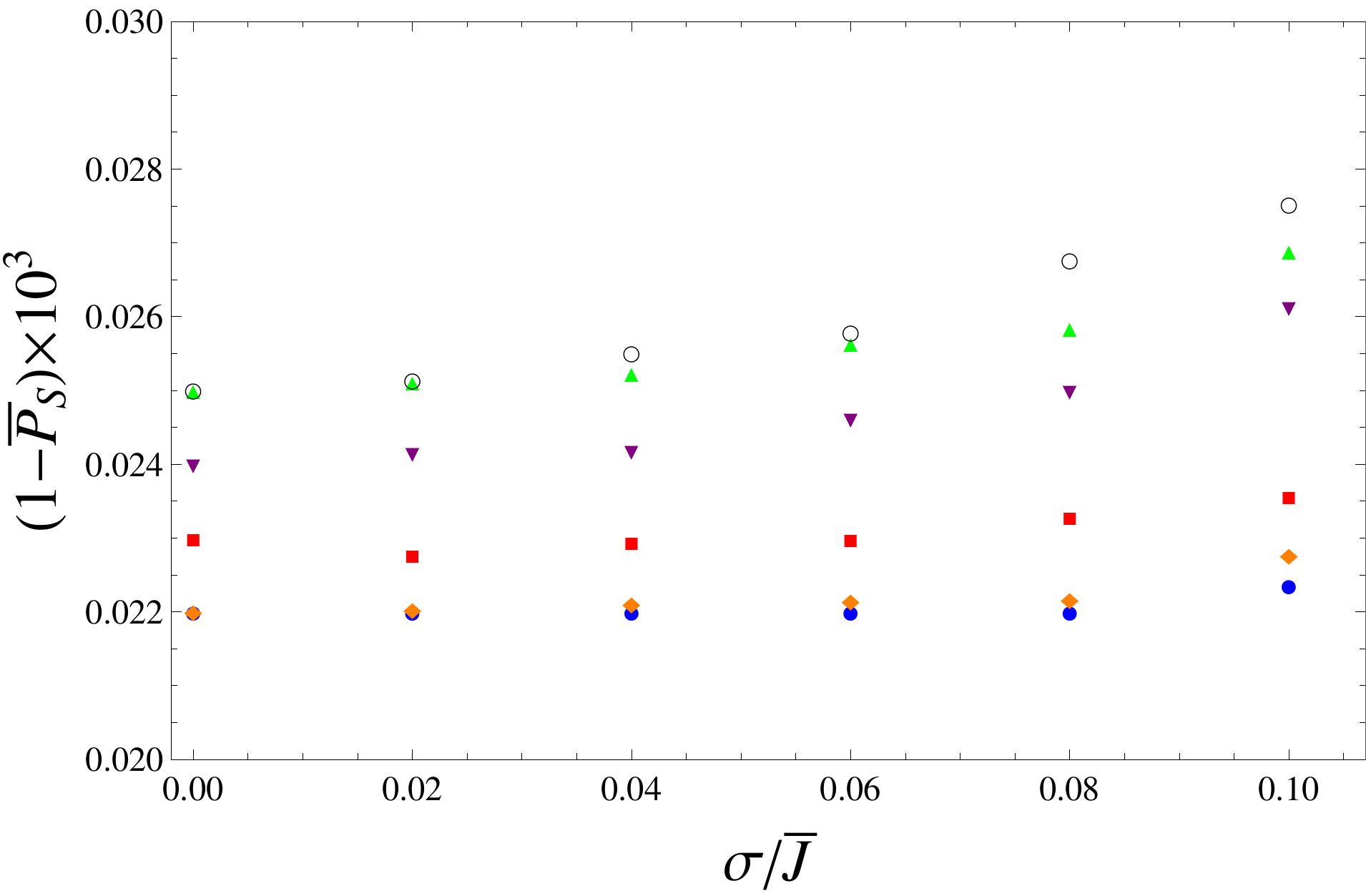}\\
 (b)\\
 \includegraphics[scale=0.32]{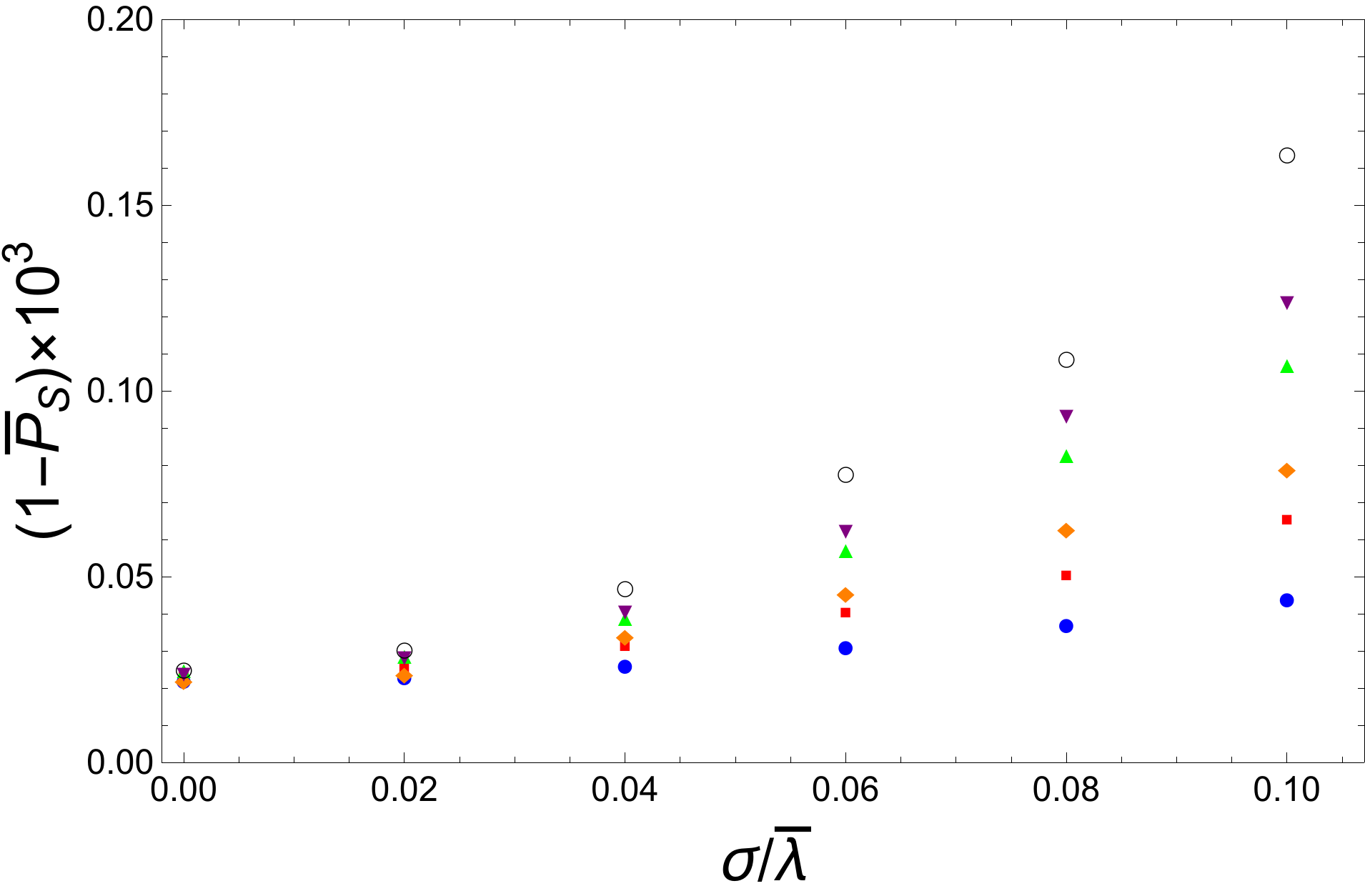}\\
 (c)
\par\end{centering}

\caption{\label{fig:PS_h&J}Average success probability for ensembles of
1024 system realizations, considering (a) $h_{i}$, (b) $J_{i,i+1}$ and (c) $\lambda_i$
as the only random variable. The probabilities are shown as a function of the the relative standard deviation of such parameters. Results for different system sizes $(N=2,3,4,5,6,8)$
are shown.}
\end{figure}

\section{Results\label{sec:Results}}

In order to quantify the computation success using the fabricated processors, we calculate the mean probability of finding the time evolved initial state into the ideal final ground state $|E_0^{(I)}(t_f)\rangle$, i.e.
\begin{equation}
\overline{P}_{S}(\sigma) \equiv  \overline{\Bigl|\langle E_0^{(I)}(t_{f})\bigl|U_{\sigma}(t_{f},0)\bigr|E_0^{(\sigma)}(0)\rangle\Bigr|^{2}},
\end{equation}
where $|E_0^{(\sigma)}(0)\rangle$ and $U_{\sigma}(t_{f},0)$ denote respectively the initial ground state and the time evolution operator of a random implementation chosen from an ensemble of Gaussian distributed physical realizations with standard deviation $\sigma$. The bar indicates the average over such an ensemble.

As already mentioned, the protocol used for each chain size $N$ was designed such that the ideal case would have at least a probability of $\sim99.9975\%$ of finding the system in its instantaneous final ground state. To maintain the same envelope function profiles, we designed them such as $\Omega(t)=\hbar\epsilon_{0}\left(1+\cos(\alpha_N t)\right)$
and $\Gamma(t)=\hbar\epsilon_{0}\left(1-\cos(\alpha_N t)\right)$, with
$\epsilon_{0}/2\pi=318.3$MHz and $\alpha_N t_{f}=\pi$ \footnote{It is worth of mention that if one introduces the dimensionless parameter $s\equiv t/t_f\in[0,1]$ using the envelope functions chosen here, the Hamiltonian Eq. \ref{eq:Hamiltonian} becomes $t_f$-independent and hence be considered having just a timescale \cite{LidarRMP}}. Naturally, since the minimum gap $\Delta_{min}$ was found to decrease with the chain size $N$, the protocol rate and hence its time duration had to be modulated such that one could reach the level of success imposed. That was done by changing the parameter $\alpha_N$ as a function of $N$ (see Fig. \ref{fig:gapminimum}). 

For characterization of the success loss due to the parameters deviations, we simulated ensembles of physical realizations, each of them having just one of the physical parameters $(\lambda,h,J)$ as a random variable. The results for ensembles of random longitudinal fields $(h_i)$, couplings $(J_{i,i+1})$ and transverse fields $(\lambda_{i})$ are shown in panels (a), (b) and (c) of Fig. \ref{fig:PS_h&J}, respectively. 

\begin{figure*}[t!]
\begin{centering}
 \includegraphics[bb=0bp 0bp 360bp 210bp,clip,scale=0.42]{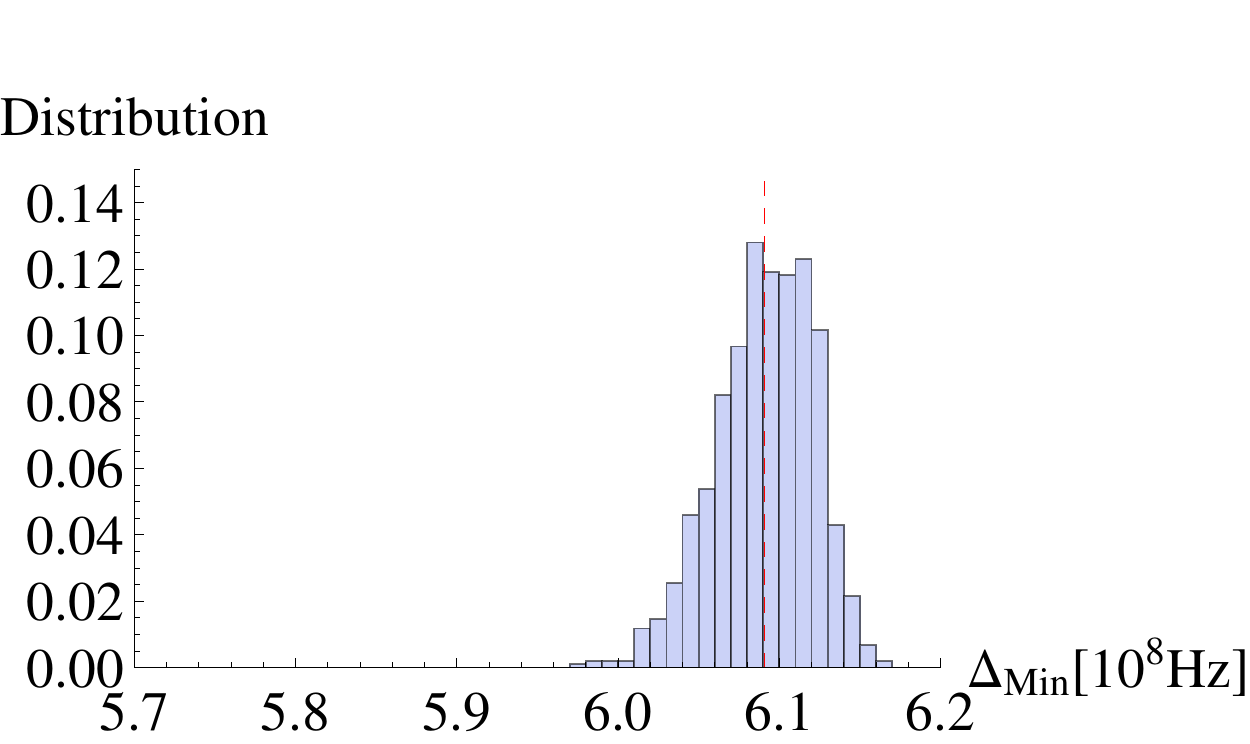}$\;$(a)$\;$\includegraphics[bb=0bp 0bp 360bp 210bp,clip,scale=0.42]{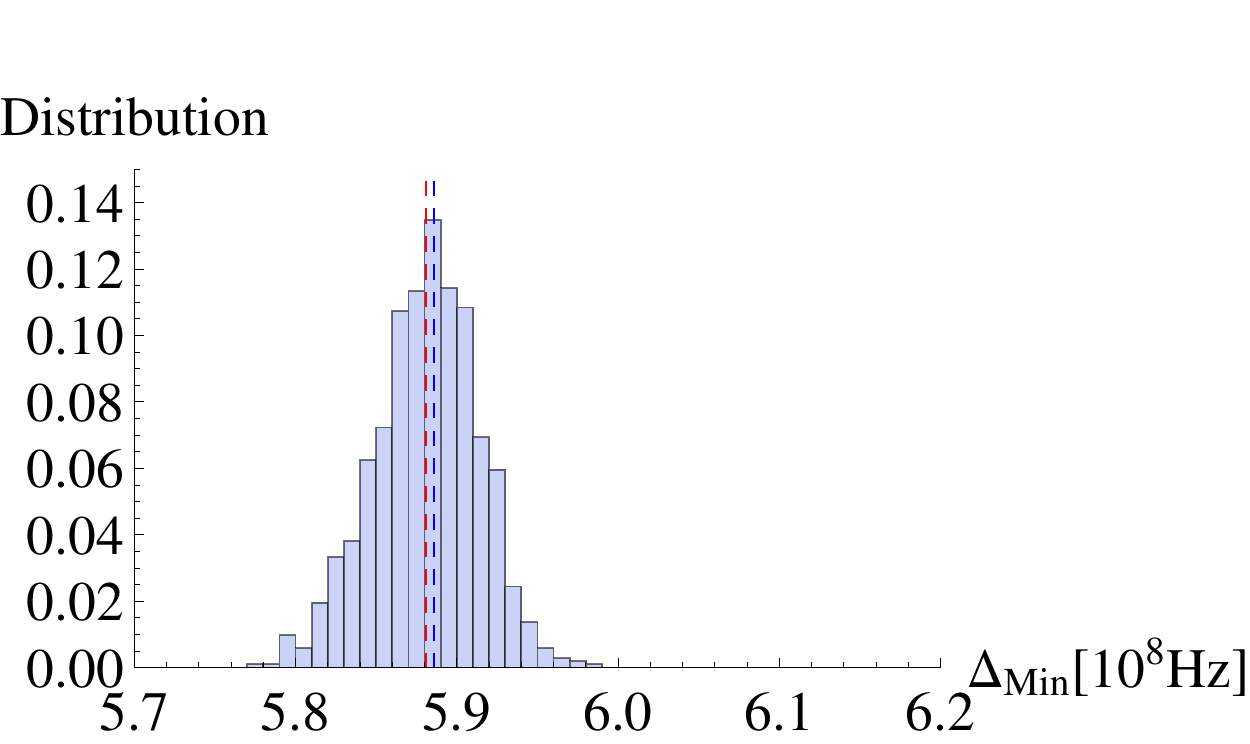}$\;$(b)$\;$\includegraphics[bb=0bp 0bp 360bp 210bp,clip,scale=0.42]{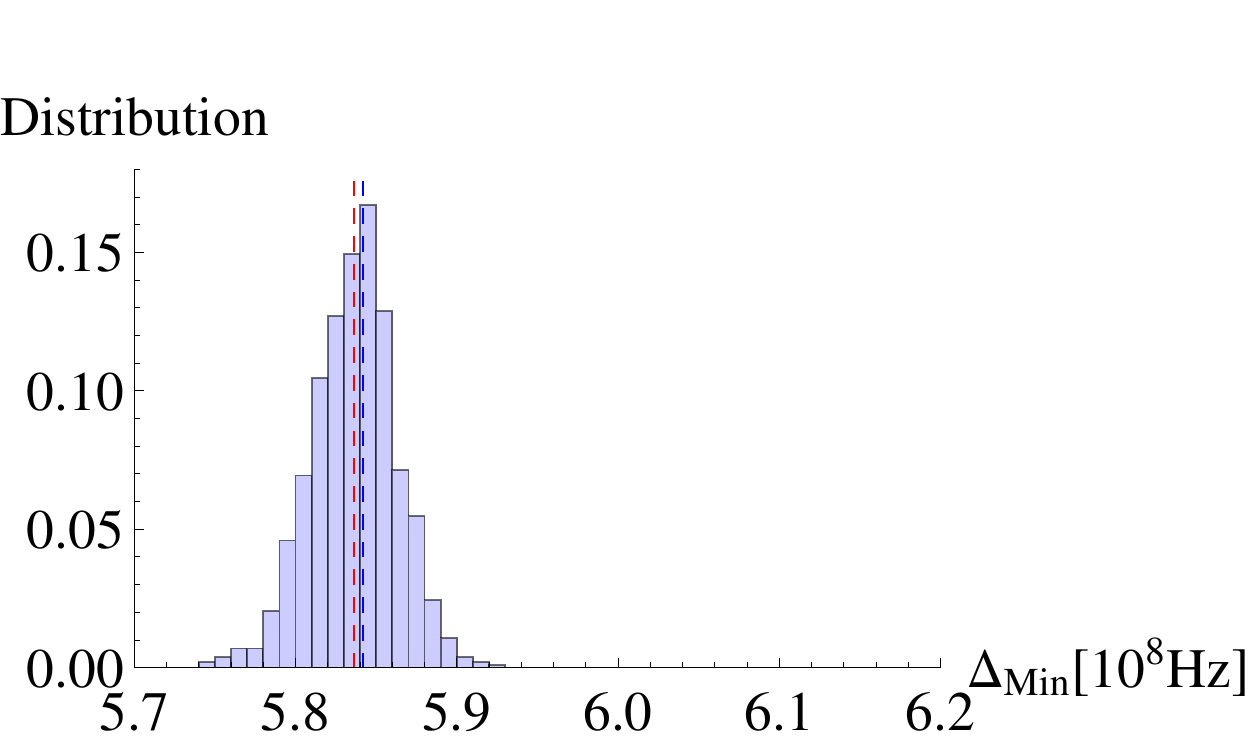}$\;$(c)\\
  \includegraphics[bb=0bp 0bp 360bp 210bp,clip,scale=0.42]{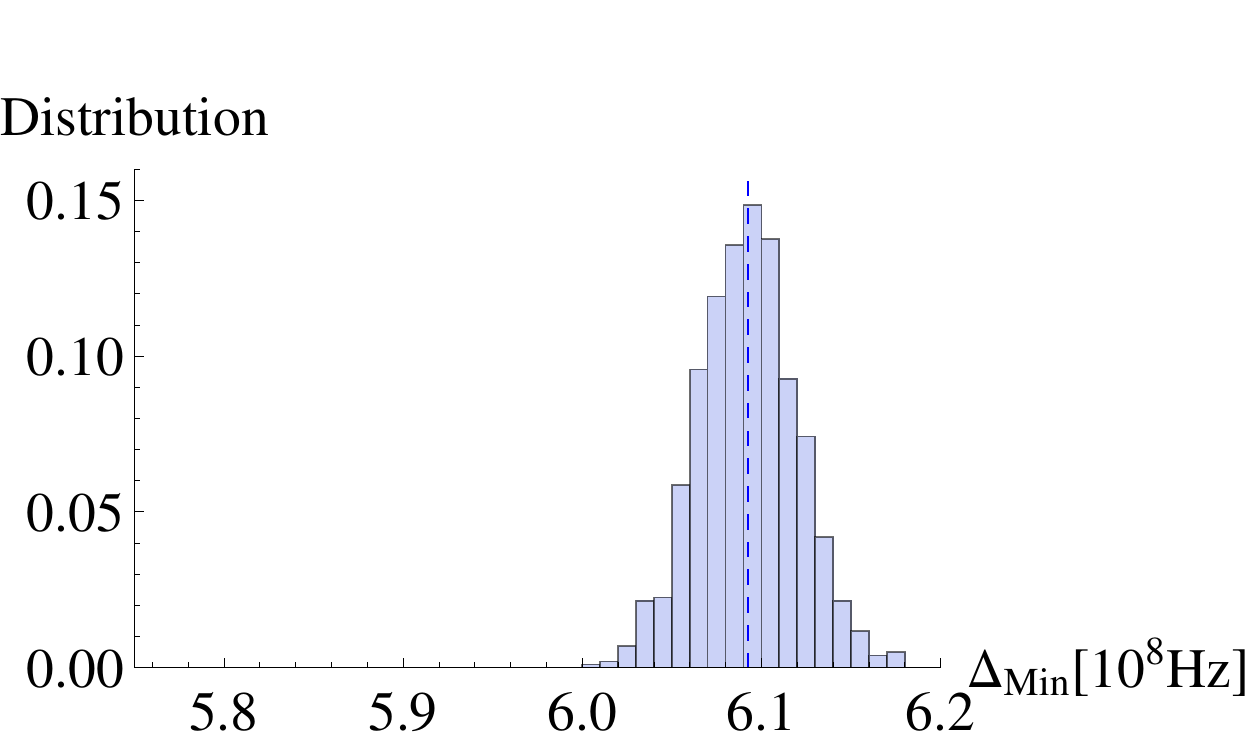}$\;$(d)$\;$\includegraphics[bb=0bp 0bp 360bp 210bp,clip,scale=0.42]{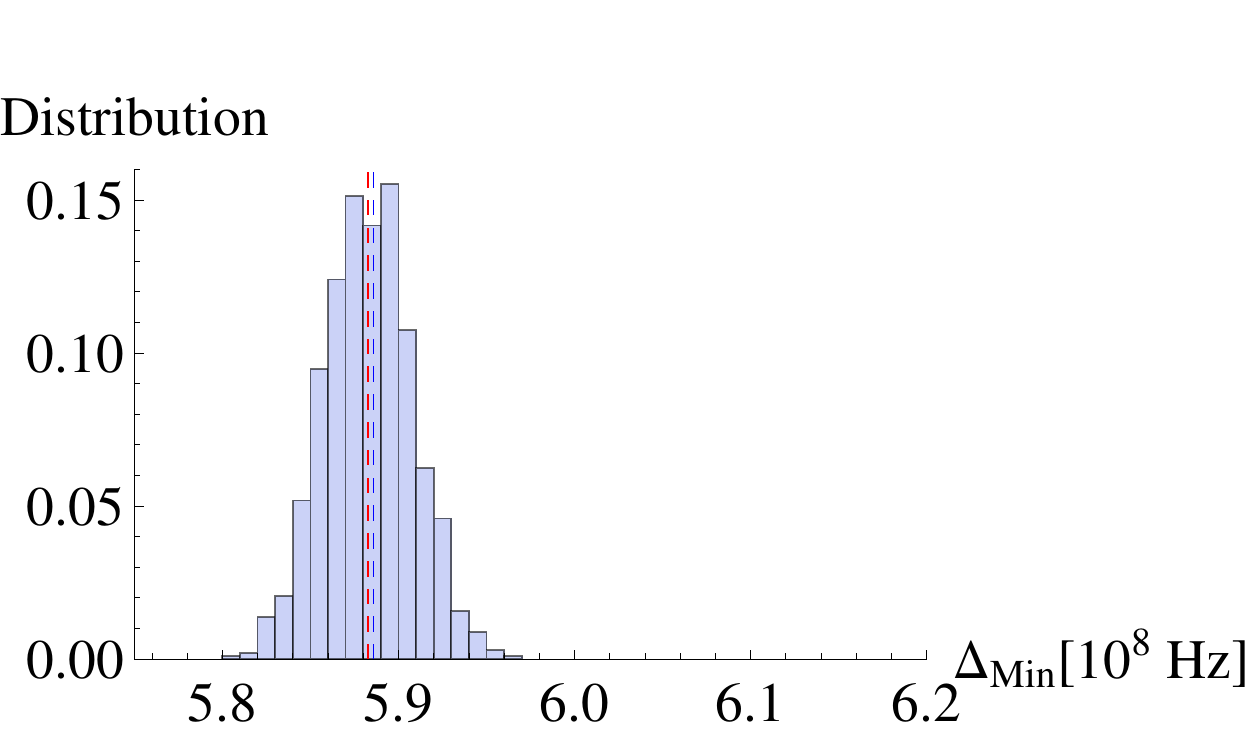}$\;$(e)$\;$\includegraphics[bb=0bp 0bp 360bp 210bp,clip,scale=0.42]{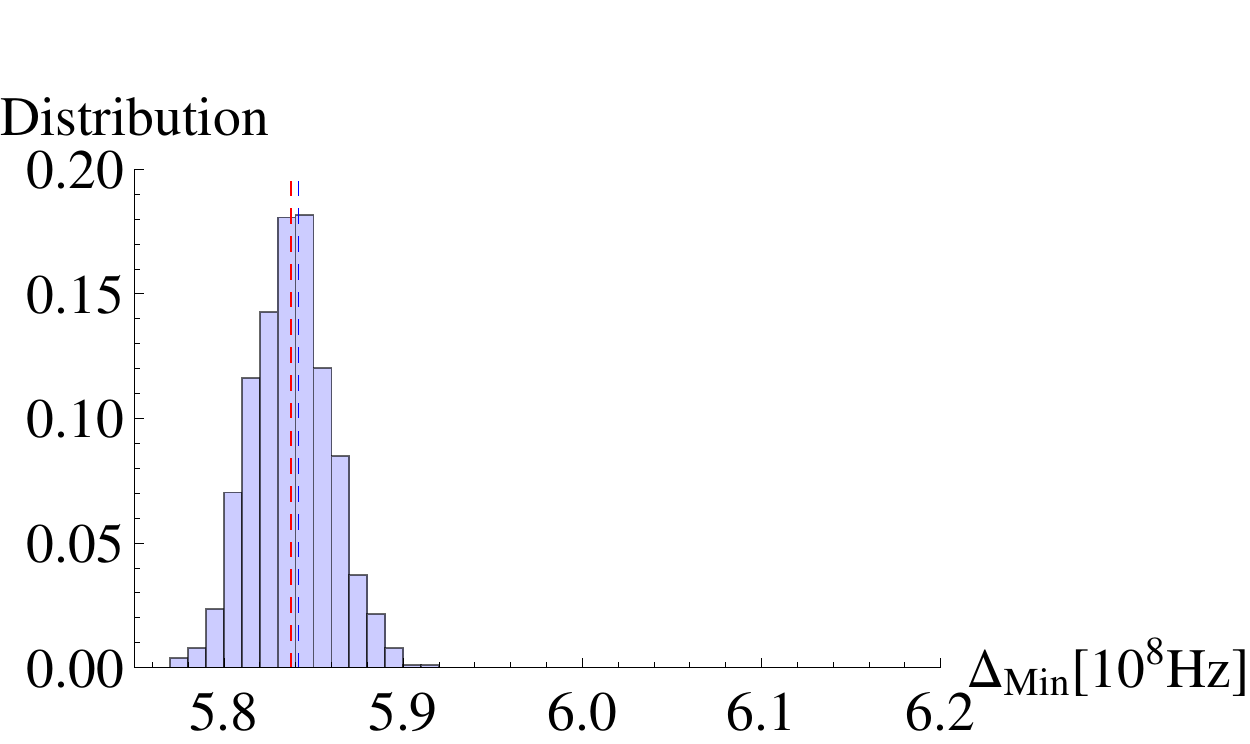}$\;$(f)\\
\includegraphics[bb=0bp 0bp 360bp 210bp,clip,scale=0.42]{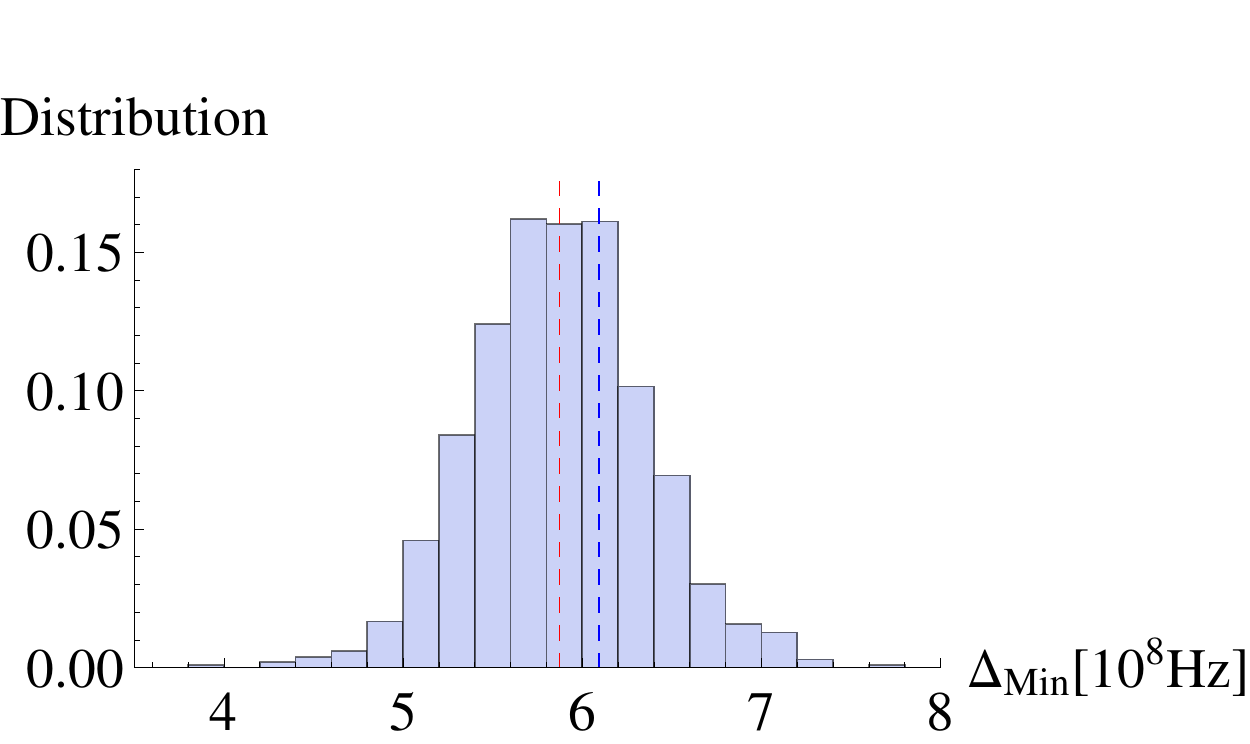}$\;$(g)$\;$\includegraphics[bb=0bp 0bp 360bp 210bp,clip,scale=0.42]{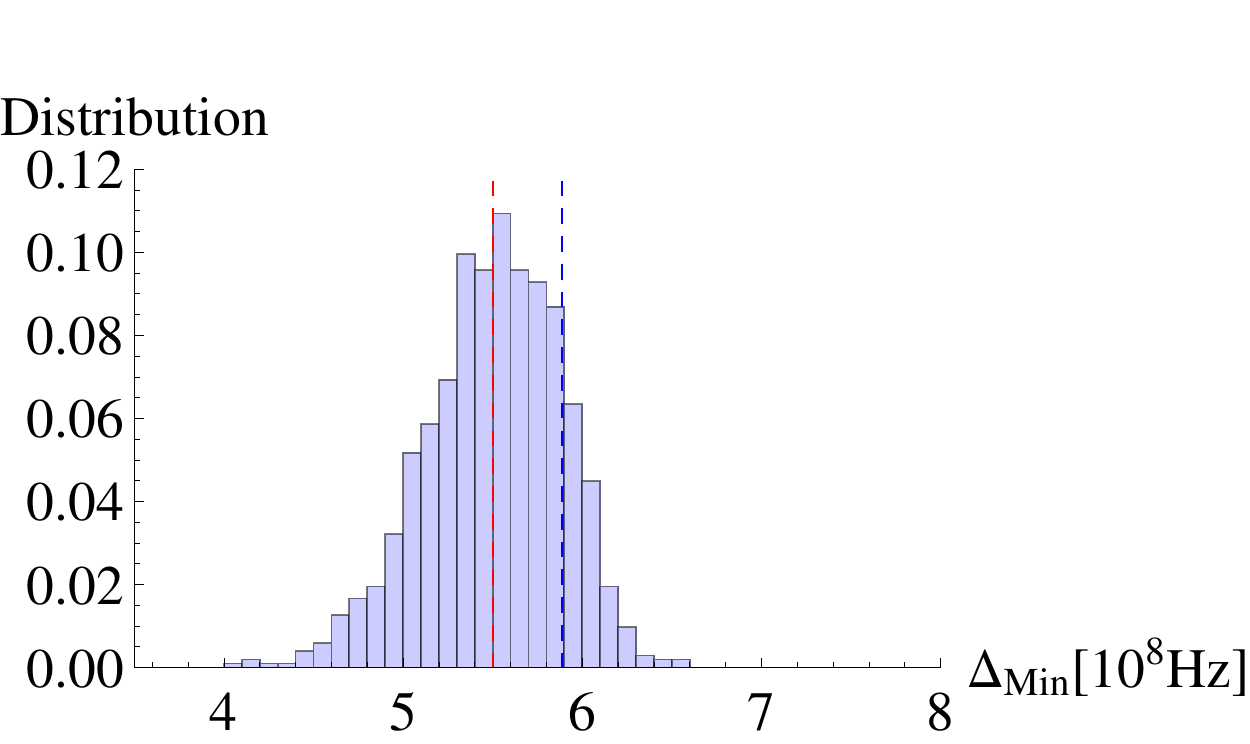}$\;$(h)$\;$\includegraphics[bb=0bp 0bp 360bp 210bp,clip,scale=0.42]{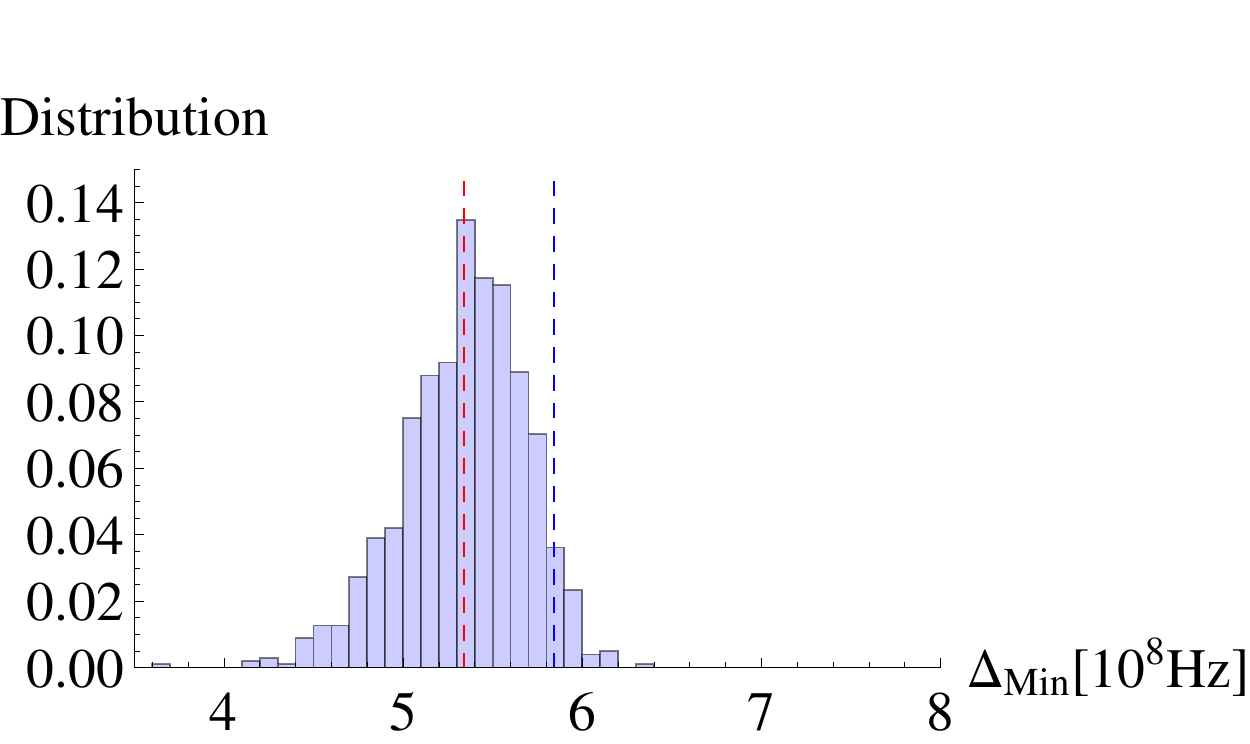}$\;$(i)
\par\end{centering}

\caption{\label{fig:Distri-min-gap}Minimum gap $\Delta_{min}$ frequency distribution
for ensembles of disordered system parameters:  $h_{i}$ (panels a-c), $J_{i,i+1}$ (panels d-f), and $\lambda_{i}$ (panels g-i). Different system sizes are shown: $N=2$ (a, d, g); $N=5$ (b, e, h); $N=8$
(c, f, i). The vertical blue line indicates the minimum gap associated with the ideal instance, while the red one shows the ensemble $\Delta_{min}$ mean value. All ensembles have 1024 system realizations, with relative standard deviation of 10\%.}
\end{figure*}

As for the disorder in $h$ and $J$, the results of panels Fig.\ref{fig:PS_h&J}(a) and (b) show that the chosen case study has the success probability subtly affected due to those deviations, even when considering relative standard deviations of the order of $10\%$. The reason for that is two fold: i) for the great majority of the instances in the ensembles, the final instantaneous ground state is the same as the one found for the ideal case; ii) for the region of appearance of the minimum gap $\Delta_{min}$, one finds that the eigenvalues of Hamiltonian Eq. (\ref{eq:Hamiltonian}) are slightly perturbed by such deviations, since the system Hamiltonian is still dominated by $H_I$, leading to an energy shift of second order on the perturbation. Indeed, when looking at the ensemble distribution of $\Delta_{min}$ registered under those conditions, Fig. \ref{fig:Distri-min-gap}(a-c) and (d-f), one finds it sharply centered in the ideal $\Delta_{min}$, even when increasing the system size $N$.

\begin{figure*}[t]
\centering
    \includegraphics[width=0.45\linewidth]{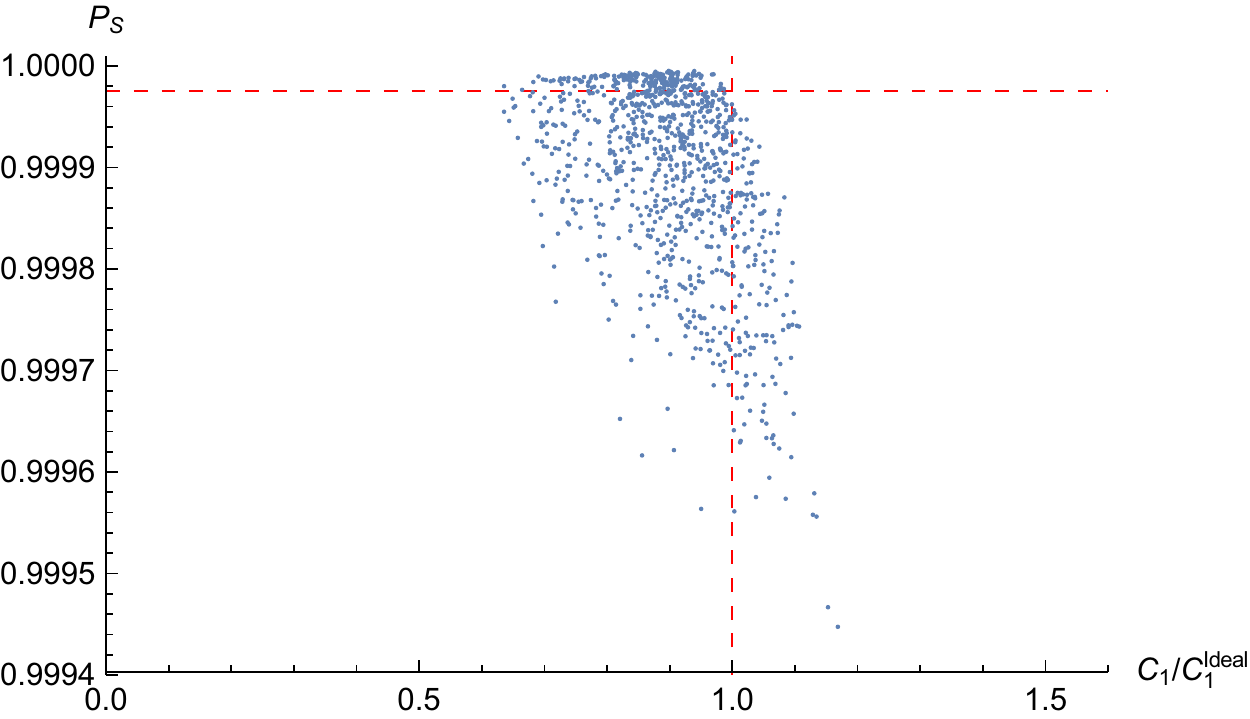}(a)\hfil
    \includegraphics[width=0.45\linewidth]{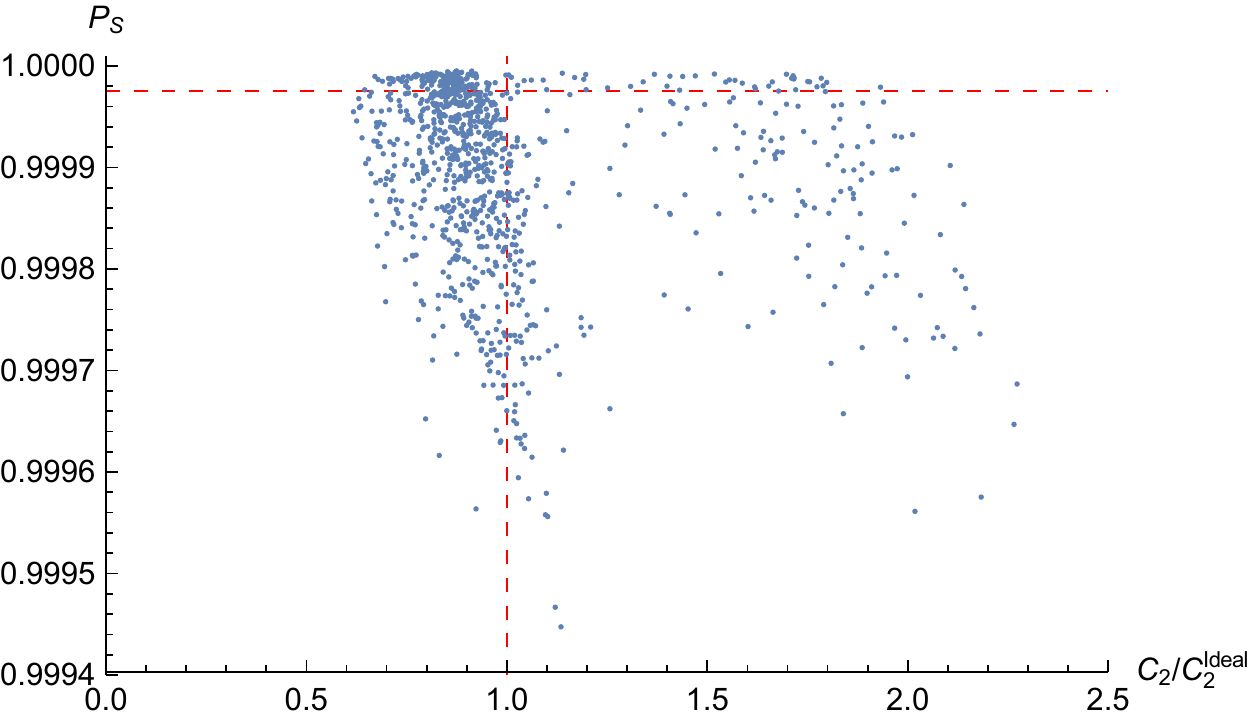}(b)\par\medskip
    \includegraphics[width=0.45\linewidth]{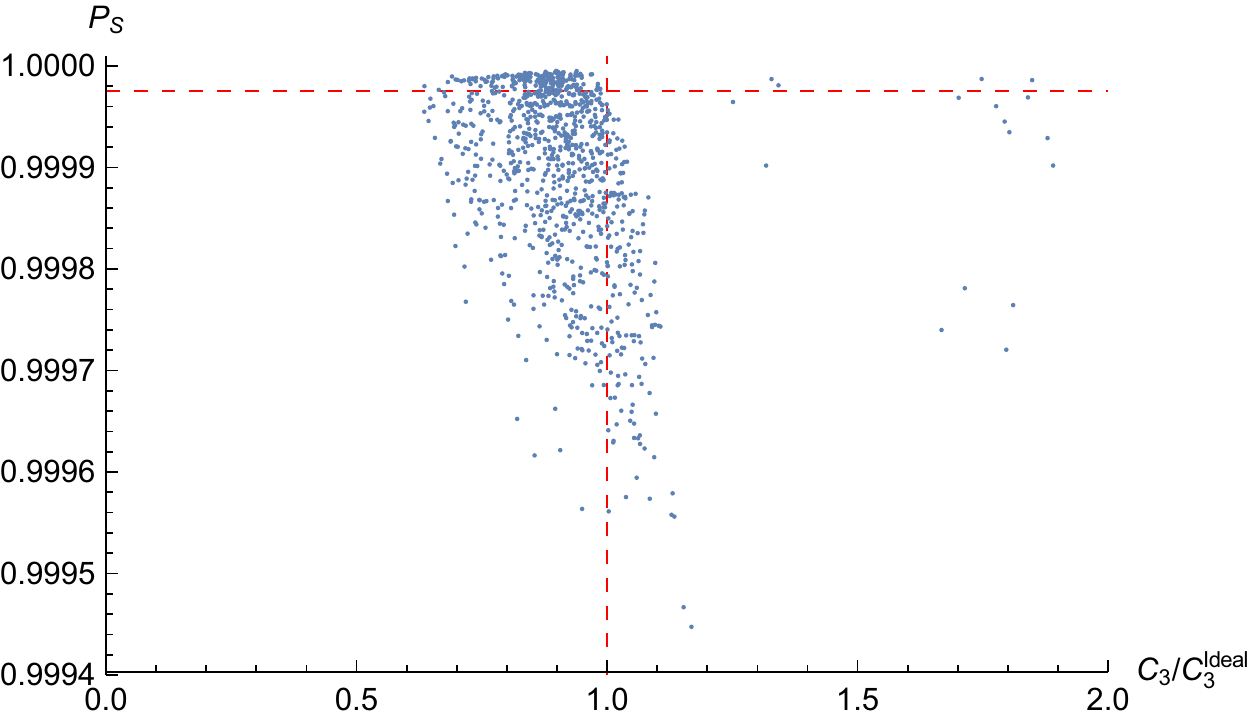}(c)\hfil
    \includegraphics[width=0.45\linewidth]{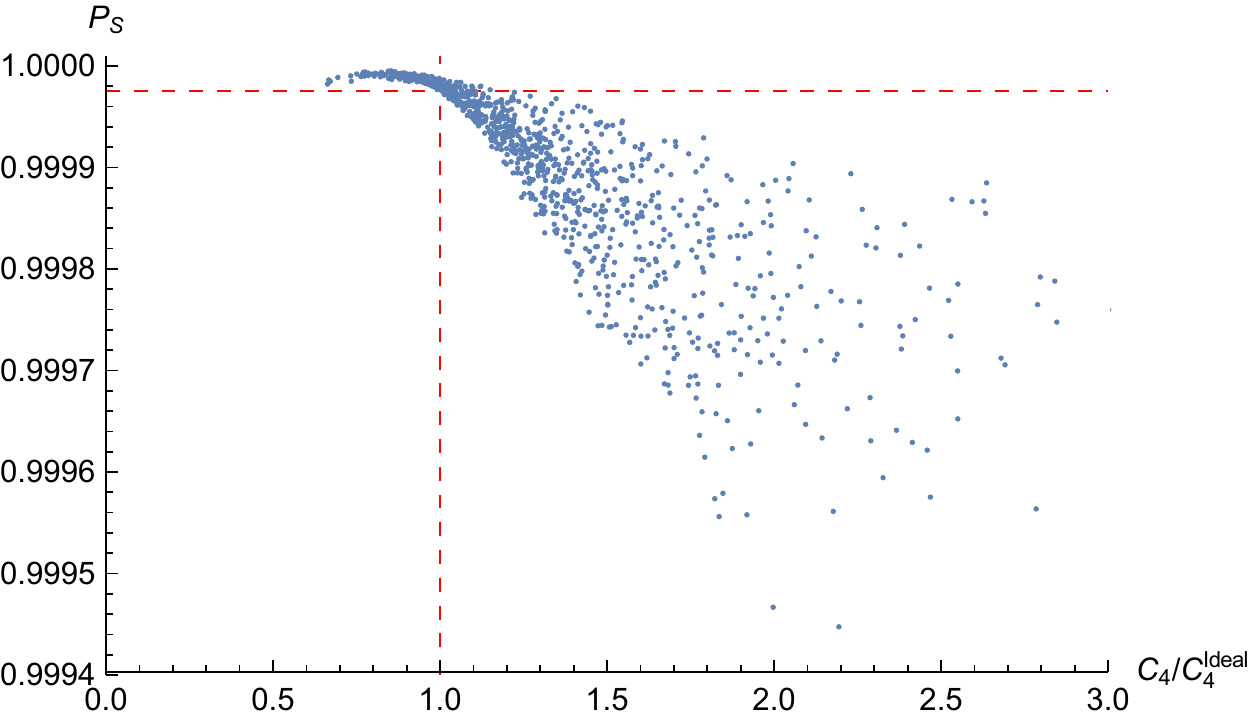}(d)
\caption{\label{fig:Conditions}Probability of success as a function of the figures of merit $C_i$, Eqs. (\ref{c1}-\ref{c4}), calculated for each one of the 1024 instances of the $10\%$ $\lambda$-disordered ensemble comprised of system size $N=5$.  The $C_i^{Ideal}$ denotes the figure of merit computed for the ideal instance. The dashed lines indicate the values related to the ideal instance.}
\end{figure*}

On the other hand, the success probability does change when considering the transverse fields $\lambda_i$ as random variables, having a noticeable dependence with the system size $N$ (see Fig.\ref{fig:PS_h&J}(c)). Nevertheless, as for the ensemble average, the probability does not degrade as one might expect by only looking at $\Delta_{min}$ distribution, Fig.\ref{fig:Distri-min-gap}(g-i), since the ensembles have much wider distributions, with just few instances presenting bigger gaps. 

Such a result calls for the attention the fact that the figure of merit for the adiabatic condition shall relate $\Delta_{min}$ with the Hamiltonian time rate $\dot{H}$. Indeed, if one writes the system time evolved state $|\psi(t)\rangle$ in terms of the instantaneous eigenenergy basis $(H(t)|E_j(t)\rangle=E_j(t)|E_j(t)\rangle)$ as $|\psi(t)\rangle=\sum_m a_m(t)e^{-\frac{i}{\hbar}\int_0^t dt' E_m(t')}|E_m(t)\rangle$, the amplitudes $a_m(t)$ are determined from a set of coupled dynamical equations obtained from the Schr\"odinger equation
\begin{equation}
\label{eq_cm}
\begin{multlined}
\frac{\partial}{\partial t}a_m(t)=-a_m(t)\langle E_m(t)|\left[\frac{\partial}{\partial t}|E_m(t)\rangle\right]\\
-\sum_{n\neq m}a_n(t)\frac{\dot{H}_{mn}(t)}{\Delta_{nm}(t)}\exp{\left(-\frac{i}{\hbar}\int_0^t dt'\Delta_{nm}(t')\right)},
\end{multlined}
\end{equation}
where $\dot{H}_{mn}(t)\equiv\langle E_m(t)|\dot{H}|E_n(t)\rangle$ and $\Delta_{nm}(t)\equiv E_n(t)-E_m(t)$. Consequently, from Eq. 8 one can envision several conditions for which the adiabatic condition $|a_m(t)|=|a_m(0)|$ could be satisfied. Actually, after the standard textbook condition \cite{MessiahBook},\begin{equation}
C_1\equiv\max_{0\leq t\leq t_{f}}\frac{\hbar\left|\dot{H}_{nm}(t)\right|}{\Delta_{nm}^2(t)}\ll 1~~\forall m\neq n,\label{c1}
\end{equation}
has been shown neither sufficient nor necessary by inspection of counter-examples \cite{PhysRevLett.93.160408,PhysRevLett.95.110407,PhysRevLett.101.060403}, which feature the system evolution having multiple timescales, a great deal of effort was put forward in order to provide a new reliable condition. As a result, diverse conditions for adiabaticity have been rigorously  formulated, but none thus far has been shown sufficient and necessary for a general case (see \cite{LidarRMP} for a timely discussion about several proposed adiabatic conditions). Since such conditions are found relying on different gap dependences, by considering the gap spread observed in our ensemble of realizations, Fig. \ref{fig:Distri-min-gap}(g-i), one could wonder if those conditions would provide the same reliability for the adiabaticity of the evolution, i.e., whether the more satisfied a condition is, the more adiabatic the evolution becomes. In order to address this point, in addition to $C_1$ (Eq. \ref{c1}), we calculated two figures of merit related with different conditions \cite{PhysRevLett.98.150402, PhysRevA.77.062114}, namely,
\begin{eqnarray}
C_2&\equiv &\hbar\int_0^{t_f}\left|\frac{d}{dt}\left(\frac{\dot{H}_{nm}}{\Delta_{nm}^2}\right)\right|dt\ll1,\label{c2}\\
C_3&\equiv&\hbar\left|\frac{\left(\dot{H}_{mn}/\Delta_{mn}\right)}{\Delta_{mn}+\delta_{nm}}\right|\ll1,\label{c3}
\end{eqnarray}
where $\delta_{nm}$ is defined as a geometric potential \cite{PhysRevA.77.062114}, related with the geometric phases accumulated during the evolution. Furthermore, a fourth figure of merit was also computed, which is related with a lower bound for the total time evolution \cite{Regev},
\begin{eqnarray}
t_f\geq\frac{10^5}{\delta^2}C_4\equiv\frac{10^5}{\delta^2}\hbar\max\left\{\frac{||H'||^3}{\Delta_{min}^4},\frac{||H'||||H''||}{\Delta_{min}^3}\right\},~\label{c4}
\end{eqnarray}
where $\delta$ is the distance between the evolved state and the corresponding instantaneous eigenstate, $||\cal{O}||$ denotes $\max_{t\in[0,t_f]} ||{\cal{O}}(t)||$, being $||\cdots||$ the usual operator norm, and $\{H',H''\}$ correspond to the first and second derivatives with respect to the dimensionless parameter $s\equiv t/t_f$. Such a set of figures of merit have been experimentally used in \cite{PhysRevLett.101.060403} for a problem of constant gap, but with multiplescales, in order to assess conditions for the adiabatic theorem, reaching the conclusion that $C_2$, $C_3$ and $C_4$ were better conditions than $C_1$ for their problem.

Our findings are shown in Fig. \ref{fig:Conditions}. Were the $C_i$ conditions quantitative figures for adiabaticity, one should find a monotonic decreasing behavior of the probability of success as $C_i$ increases. However, as depicted in the panels of Fig. \ref{fig:Conditions}(a-d), such a behavior does not happen for our problem. Actually, for $C_1$, $C_2$ and $C_3$, Fig. \ref{fig:Conditions}(a-c), the results reveal that there is no regime were those conditions could be taken as quantitative measures for adiabaticity. Therefore, lowering such $C_i$'s does not necessarily mean improving the probability of success. As for the condition $C_4$, one finds that it becomes a more reliable quantity as its value decreases, but such a behavior seems only to happen close to the saturation value, i.e., $P_S=1$.

\section{Conclusions}

In summary, we have examined numerical simulations to investigate the performance of a quantum adiabatic processor using as physical resources superconducting flux qubits, calculating the probability of reaching an ideal final state, under fabrication errors of these devices. 

We have demonstrated the robustness of the model (adiabatic quantum computation)
against errors of fabrication and manipulation when is considered an ensemble of disordered instances in both $J_{i,j}$ and $h_{i}$. Nevertheless, a fragility was found when the disorder in the transversal fields $\lambda_i$ was considered, which is directly related with the physical parameters, i.e., $C$, $L$ and $E_J$, of each superconducting qubit. 

In addition, the problem chosen here allowed us to eliminate the source of error due to the change of the final ground state, giving a clear view of the contributions of the dynamical errors generated by deviations from the ideal eigenenergy dynamics. Under such conditions, it was found that the degradation of the probability of success could not directly be related with the variations of the minimum gap observed for each instance. By analyzing several proposed conditions for adiabaticity, we have found that those conditions could not be used to quantify the adiabaticity of the protocol. Our results indicate that seeking for a better compliance with adiabatic conditions does not necessarily lead to computation improvements, giving evidences that such an approach for adiabatic quantum computation may not be optimal.
\section*{Acknowledgments}

We gratefully acknowledge financial support from Conselho Nacional
de Desenvolvimento Cient\'{i}fico e Tecnol\'ogico (CNPq). FB is supported by the Instituto Nacional de Ci\^encia e Tecnologia - Informa\c{c}\~ao Qu\^antica (INCT-IQ).

%\addcontentsline{toc}{section}{\refname}
\bibliographystyle{apsrev4-1}
\bibliography{References_paper}

\end{document}